\documentclass[12pt]{emulateapj}

\renewcommand{\min}{\mbox{$^m$}}

\renewcommand{\deg}{\mbox{$^{\circ}$}}

\def\deg{\ifmmode^\circ\else$^\circ$\fi}    

\def\hper{\ifmmode \rlap.^{h}\else $\rlap{.}^h$\fi} 
\def\sper{\ifmmode \rlap.^{s}\else $\rlap{.}^s$\fi}    

\def\deg{${}^\circ$}
\def\min{${}^{\prime}$}

\def\vmi{\hbox{\it V--I\/}}
\def\bmv{\hbox{\it B--V\/}}

\def\today{\number\year\space \ifcase\month\or
January\or February\or March\or April\or May\or June\or
July\or August\or September\or October\or November\or December\fi
\space\number\day}
\def\now{\number\year\space \ifcase\month\or
January\or February\or March\or April\or May\or June\or
July\or August\or September\or October\or November\or December\fi
\space\number\day .\number\time}

\shorttitle{On the radial extent of the dwarf irregular galaxy IC10} 
\shortauthors{Sanna et al.}

\begin{document}

\title{On the radial extent of the dwarf irregular galaxy IC10\altaffilmark{1}} 

\author{
N.\ Sanna\altaffilmark{2,3},
G.\ Bono\altaffilmark{3,4},
P.\ B.\ Stetson\altaffilmark{5},
I.\ Ferraro\altaffilmark{4},
M.\ Monelli\altaffilmark{6},
M.\ Nonino\altaffilmark{7}, 
P.\ G.\ Prada Moroni\altaffilmark{8,9},
R.\ Bresolin\altaffilmark{10},
R.\ Buonanno\altaffilmark{3,11},
F.\ Caputo\altaffilmark{4},
M.\ Cignoni\altaffilmark{12,13},
S.\ Degl'Innocenti\altaffilmark{8,9}, 
G.\ Iannicola\altaffilmark{4},
N.\ Matsunaga\altaffilmark{14},
A.\ Pietrinferni\altaffilmark{15},
M.\ Romaniello\altaffilmark{16},
J.\ Storm\altaffilmark{17}, and  
A.\ R. \ Walker\altaffilmark{18}
}

\altaffiltext{1}{
This research used the facilities of the Canadian Astronomy Data Centre operated 
by the National Research Council of Canada with the support of the Canadian Space 
Agency. This research is based in part on data collected at Subaru Telescope, 
which is operated by the National Astronomical Observatory of Japan.  
}
\altaffiltext{2}{University of Virginia, 530 McCormick Road 22903 Charlottesville, 
VA, USA; ns2as@virginia.edu}
\altaffiltext{3}{Univ. Roma Tor Vergata, via della Ricerca Scientifica 1, 00133 Rome, Italy}
\altaffiltext{4}{INAF--OAR, via Frascati 33, Monte Porzio Catone, Rome, Italy}
\altaffiltext{5}{DAO--HIA, NRC, 5071 West Saanich Road, Victoria, BC V9E 2E7, Canada}
\altaffiltext{6}{IAC, Calle Via Lactea, E38200 La Laguna, Tenerife, Spain}
\altaffiltext{7}{INAF--OAT, via G.B. Tiepolo 11, 40131 Trieste, Italy}
\altaffiltext{8}{Univ. Pisa, Largo B. Pontecorvo 2, 56127 Pisa, Italy}
\altaffiltext{9}{INFN, Sez. Pisa, via E. Fermi 2, 56127 Pisa, Italy}
\altaffiltext{10}{Institute for Astronomy, 2680 Woodlawn Drive, Honolulu, HI 96822, USA} 
\altaffiltext{11}{ASI--Science Data Center, ASDC c/o ESRIN, via G. Galilei, 00044 Frascati, Italy}
\altaffiltext{12}{Dipartimento di Astronomia, Universit\`a di Bologna, via Ranzani 1, 40127 Bologna, Italy}
\altaffiltext{13}{INAF--OAB, via Ranzani 1, 40127 Bologna, Italy}
\altaffiltext{14}{Institute of Astronomy, University of Tokyo, 2-21-1 Osawa, Mitaka, Tokyo 181-0015, Japan}
\altaffiltext{15}{INAF--OACTe, via M. Maggini, 64100 Teramo, Italy}
\altaffiltext{16}{ESO, Karl-Schwarzschild-Str. 2, 85748 Garching bei Munchen, Germany}
\altaffiltext{17}{AIP, An der Sternwarte 16, D-14482 Potsdam, Germany}
\altaffiltext{18}{NOAO--CTIO, Casilla 603, La Serena, Chile}

\date{\centering drafted \today\ / Received / Accepted }

\begin{abstract}
We present new deep and accurate space (Advanced Camera for Surveys -- Wide Field 
Planetary Camera 2 at the Hubble Space Telescope) and ground-based (Suprime-Cam at Subaru
Telescope, Mega-Cam at Canada-France-Hawaii Telescope) photometric and astrometric data 
for the Local Group dwarf irregular IC10. 
We confirm the significant decrease of the young stellar population when moving from the 
center toward the outermost regions. We find that the tidal radius of IC10 is significantly 
larger than previous estimates of $r_t \lesssim$ 10\min.
By using the $I$,\vmi\ Color Magnitude Diagram based on the Suprime-Cam data we detect sizable
samples of red giant (RG) stars up to radial distances of 18-23$'$ from the galactic center. 
The ratio between observed star counts (Mega-Cam data) across the tip of the RG branch and star
counts predicted by Galactic models indicate a star count excess at least at a 3$\sigma$ level 
up to 34-42\min\ from the center. This finding supports the hypothesis that the huge 
H{\footnotesize{I}} cloud covering more than one degree across the galaxy is 
associated with IC10  \citep{huchtmeier79,cohen79}. 
 We also provide new estimates of the total luminosity
($L_V\sim$9$\times$$10^7$ $L_\odot$, $M_V$$\sim$-15.1 mag) that agrees with 
similar estimates available in the literature. If we restrict to the regions 
where rotational velocity measurements are available (r$\approx$13$'$), 
we find a mass-to-light ratio ($\sim$10 $M_\odot$/$L_\odot$)
that is at least one order of magnitude larger than previous estimates. The
new estimate should be cautiously treated, since it is based on a minimal
fraction of the body of the galaxy. 
\end{abstract}


\maketitle

\section{Introduction}
Photometric investigations of the stellar populations in Local Group (LG) 
dwarf galaxies provide firm constraints on cosmological parameters and 
the unique opportunity to investigate galaxy formation models 
\citep{mateo98,tolstoy09,wyse10}.
In this context dwarf irregulars (dIs) play a key role, since we still 
lack firm empirical and theoretical constraints concerning their evolution and 
the possible transition into dwarf spheroidal galaxies 
\citep{bekki08,woo08,kormendy09}.  
Although the number of dwarf galaxies known in the LG is rapidly growing in 
the last few years, current statistics indicate that the dIs are at least 
one quarter of LG galaxies \citep{mcconnachie08,sanna09}. 

Among the dIs of the LG, IC10 is an interesting system, since it underwent  
strong star formation activity during the last half billion years and it is 
considered the only LG analog of starburst galaxies. 
Even though IC10 has been the 
subject of several investigations ranging from the radio \citep{wilcots98} 
to the near-infrared \citep[NIR,][]{vacca07}, to the UV \citep{hunter01,richer01}, 
and to the X-ray \citep{wang05}, its structural 
parameters and in particular its radial extent are poorly defined. 
\citet{massey95} found that the major axis of IC10 is $\sim 7'$. 
A similar diameter ($\sim 6'$) was found by \citet{jarrett03} 
using the isophotal radii from 2MASS NIR images. 
More recently \citet{tikhonov09},  using both ground-based and 
space images, suggested that the extent of the thick disk along the minor axis 
is $\approx$10.5$'$. It has also been suggested by \citet{demers04}, 
using asymptotic giant branch and red giant branch (RGB) stars,  
that IC10 should have a halo of $\sim 30'$ diameter. 
On the other hand, radio measurements by \citet[hereinafter H79]{huchtmeier79} 
indicated that IC10 has a huge envelope of neutral hydrogen extending over more 
than 1 square degree ($62 '\times 80'$) across the sky. 

We are also facing a significant uncertainty in the total mass of IC10. By using 
H{\footnotesize{I}} regions H79 found $M_{tot}\sim1.8$x$10^9M_{\odot}$, assuming 
a distance of 1 Mpc and a Holmberg diameter of $\sim 10'$.  
Also, \citet[hereinafter SS89]{shostak89}, using high resolution maps of 
H{\footnotesize{I}} regions, measured an inclination of 45\deg~ and a maximum 
in the rotation curve of 30 km s$^{-1}$ (42 km s$^{-1}$ deprojected) and the 
same Holmberg diameter (deprojected angular diameter $\sim 13'$), from which they found 
$M_{tot}\sim1$x$10^9M_{\odot}$. More recently, \citet{vandenbergh00}, following H79,
but assuming a smaller distance \citep[660 kpc,][]{sakai99}, found $M_{tot}\sim 6$x$10^8M_{\odot}$.

\section{Photometric data}
The Hubble Space Telescope (HST) data were collected with the 
Advanced Camera for Surveys (ACS, pointings $\alpha$,$\beta$) 
and with the 
Wide Field Planetary Camera 2 (WFPC2, pointings $\gamma$,$\delta$). 
Data from pointings $\alpha$, $\beta$ and $\gamma$  were already presented 
\citep{sanna08,sanna09}. Pointing $\delta$ includes 24 $F555W$-band 
and 24 $F814W$-band images of 500 s each. This pointing is located $\sim 3'$ 
NE of the galaxy center; it is outside the disk identified by 
\citet[see the red ellipse in Fig.~1]{jarrett03} and inside the thick 
disk identified by \citet{tikhonov09}. 
The ground-based data were collected with the prime focus camera (Suprime-Cam, 
pointing $\epsilon$) on the Subaru telescope and with Mega-Cam on the CFHT 
(pointing $\zeta$). Pointing $\epsilon$ (see Fig.~1)
is located across the galaxy center and includes both shallow 
(3$V$, 3$R$, 3$I$; 3$\times$60 s per band) and deep (12$V$, 12$\times$480 s; 
13$R$, 13$\times$360 s; 23$I$, 23$\times$240 s) images. 
The pointing $\zeta$ (see Fig.~1) is also located across the galaxy center
and includes 3g (3$\times$700 s) and 3i-band (3$\times$400 s) images.

\begin{figure*}[htbp]
\begin{center}
\includegraphics[height=0.45\textheight,width=0.65\textwidth]{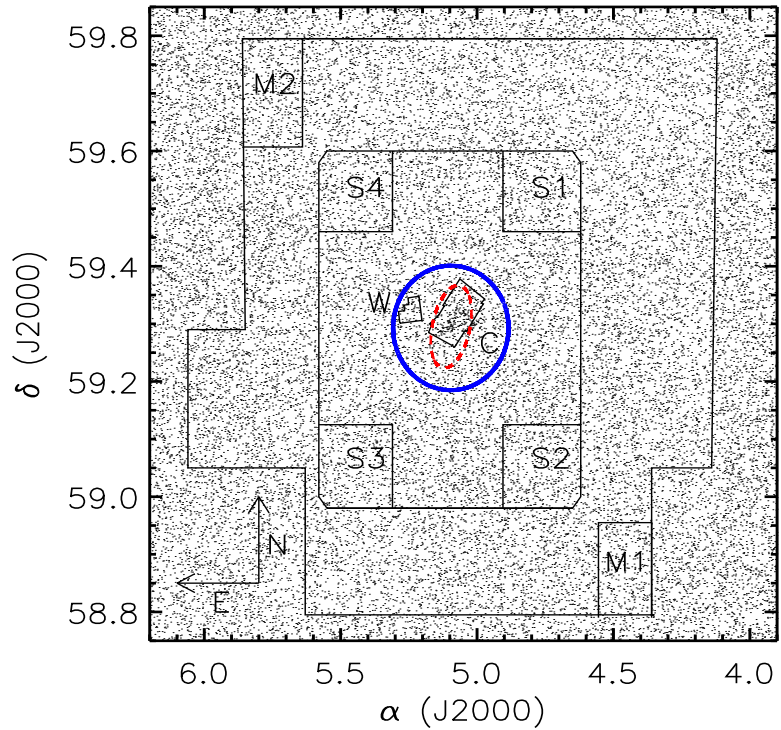}
\caption{Location of ground-based and space pointings. 
The background is a reference image of IC10 based on randomly selected stars 
from the Digitized Sky Survey (DSS) catalog. 
The polygon labeled "C" shows the sky area covered by pointings $\alpha$ (ACS@HST) 
and $\gamma$ (WFPC2@HST), the polygon ``W" the pointing $\delta$ (WFPC2@HST). 
The large rectangular polygon shows the pointing $\epsilon$ (Suprime-Cam@Subaru), 
and the largest irregular polygon the pointing $\zeta$ (Mega-Cam@CFHT). 
The boxes labeled ``S1", ``S2", ``S3", ``S4" and ``M1", ``M2" display the regions adopted 
for the CMDs. The red ellipse displays the estimate of major and minor axis 
according to \citep{jarrett03}. The blue circle has a diameter of 13$'$.}  
\end{center}
\end{figure*}

Photometry on individual images was performed with {\tt DAOPHOT IV/ALLSTAR} 
\citep{stetson87}. The 786 ground-based images were simultaneously reduced 
with {\tt ALLFRAME} \citep{stetson94}; the same applies to the 392 space images.  
We ended up with a catalog including $\sim 1,200,000$ stars 
with at least one measurement in two different bands. 
The ground-based data were transformed into the  Johnson-Kron-Cousins system 
using the standard stars provided by \citet{landolt83,landolt92} to calibrate 
local standards.  
The typical accuracy is 0.04 for $I$ and 0.05 mag for the $V$ band. 
Some external chips of the Mega-Cam include a limited number of local standards 
and they were not included in the final calibrated catalog. 
To provide a homogeneous photometric catalog the ACS and the WFPC2 were 
transformed into the $V,I$ Johnson-Kron-Cousins system using prescriptions by 
\citet{sirianni05}. The typical accuracy is of a few hundredths of 
magnitude in both $V$ and $I$. The conclusions of this investigation are not 
affected by the precision of the absolute zero-points.  

\section{Results and Discussion}
The ground-based data cover a sky area 
of $\sim 1^{\circ} \times 1^{\circ}$ while the high angular resolution of HST 
images allowed us to perform accurate photometry in the innermost crowded 
regions.  We selected eight different regions, the region ``C" is located 
across the galaxy center and includes data of pointings 
$\alpha$(ACS@HST) and $\gamma$ (WFPC2@HST), while the region ``W" is located 
$\sim3'$ from the center and includes the data of the pointing $\delta$ (WFPC2@HST). 
Regions ``S1", ``S2", ``S3", and ``S4" cover the corners of the Suprime-Cam data 
(pointing $\epsilon$) 
and are located at $\sim18'$ from the galaxy center, while regions ``M1" and ``M2" are two 
regions of the Mega-Cam data (pointing $\zeta$) located at $\sim30'$ from the galaxy center 
in the SW and NE directions, respectively.
Fig.~2 shows the $I$,\vmi\ Color-Magnitude Diagrams (CMDs) of the selected regions.  
Data plotted in this figure show several interesting features. 

{\em i)} The photometry based on HST data is deep and very accurate, and indeed the CMDs 
(regions ``C" and ``W") reach limiting magnitudes of $I\sim25.5-26$ and $V\sim26.5-27$ mag.	
The same outcome applies to the Subaru data, and indeed the CMDs reach limiting magnitudes of 
$I\sim25$ and $V\sim26$~mag. The CMDs based on CFHT data are shallower with limiting magnitudes 
$I\sim22.5$ and $V\sim24.5$ mag.     

{\em ii)} Young MS stars ($18 \le I \le 25.5$, $1 \le \vmi \le 1.5$ mag) show a strong radial gradient. 
Their number decreases rapidly when moving from the center to the outermost galaxy regions. A handful 
of them are visible in region ``W", while the blue objects of region ``S" might be field galaxies.     

\begin{figure*}[htbp]
\begin{center}
\includegraphics[height=0.65\textheight,width=0.85\textwidth]{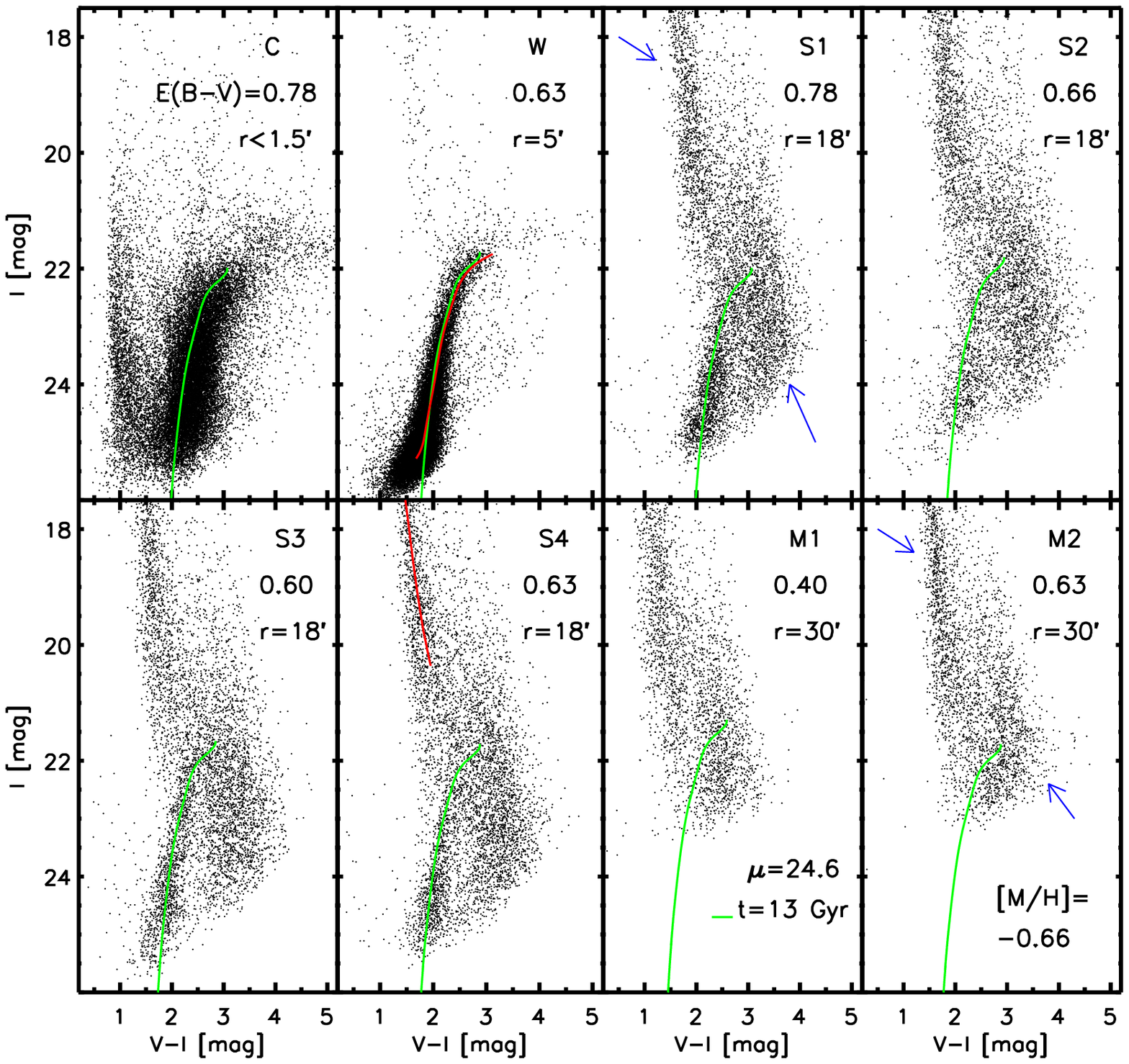}
\caption{I,V-I CMDs for eight regions located at different radial distances (see labeled values and Fig.~1). 
Blue young, main sequence stars decrease quite rapidly when moving from the center to the outermost regions. 
The old RGB stars are ubiquitous in HST and in Subaru data. The reddening also changes when moving across 
the different regions (see labeled values). The green line shows a $\alpha$-enhanced isochrones of 13 Gyr, 
at fixed chemical composition ([M/H]=-0.66 dex, Y=0.25) from the BaSTI database. The blue arrows display the 
location of contaminating field stars, while the red lines plotted in panel ``W" and ``S4" display the 
ridgeline used to determine the reddening.}
\end{center}
\end{figure*}

{\em iii)} The different apparent colors of RGB stars when moving from the center to the outermost 
regions further support the occurrence of differential reddening. We estimated the reddening 
of region ``W" using the same approach adopted in \citet{sanna08}. The ridgeline of the RGB 
in this field was adopted to estimate the reddening in the regions covered by Subaru data. 
The ridgeline of the contaminating blue field stars located in region ``S4" was used to 
estimate the reddening in the regions covered by CFHT data. We found 
that the reddening is higher along the semi-major axis ($E(\bmv)=0.78\pm0.10$ mag) and attains an 
almost constant value ($E(\bmv)=0.63\pm0.10$ mag) in the regions covered by the Subaru data external 
to the HST data. In the regions covered by the CFHT data external to the Subaru data the 
reddening attains either similar ($E(\bmv)=0.63\pm0.10$ mag) or smaller ($E(\bmv)=0.40\pm0.10$ mag) 
values in the SW direction (Sanna et al.\ 2010, in preparation). The presence of a 
significant reddening variation up to 20$'$ from the center supports previous investigations 
based on the radial distribution of H{\footnotesize{I}} regions (SS89; \citealt{wilcots98}). 

{\em iv)} Old RGB stars are ubiquitous and they can be easily identified in HST and Subaru 
CMDs. To properly identify RGB stars we also plotted an $\alpha$-enhanced isochrone 
(green line) of  13 Gyr at fixed metal content (total metallicity, [M/H]=-0.66 dex; helium content, 
Y=0.251) from the BaSTI data base \citep{pietrinferni06}\footnote{See also http://www.oa-teramo.inaf.it/BASTI}. 
We also adopted the same true distance modulus ($\mu$=24.60$\pm$0.15 mag, \citealt{sanna08}) and 
individual reddening estimates. The comparison between theory and observations 
further supports the evidence that stars located in the range $21 \lesssim I \lesssim 26$ and 
$1 \lesssim \vmi \lesssim 2.5$ mag are bona fide old and intermediate-age RGB stars. 
The CMDs based on these data show also strong contamination by field stars (see the 
region ``S1").    

{\em v)} The CMDs based on CFHT data show the same field contamination as the Subaru data,
and probably a small overdensity of stars in the region across the tip of the RGB 
(TRGB, $21 \lesssim I \lesssim 22$, $2 \lesssim \vmi \lesssim 3.5$ mag).   

The above results indicate that the radial extent of IC10 has been significantly 
underestimated, and indeed according to the Subaru data the diameter is at least 
of the order of 36-46$'$ and probably larger than one degree according to the CFHT data. 

To further constrain the radial extent of IC10, we decided to compare the observed 
star counts with star counts of foreground field stars predicted by Milky Way (MW)  
models. This approach presents several advantages when compared with the method 
based on the statistical subtraction of an external control field. 
{\em i)} It is not affected by reddening differences between the galactic and the 
control field. {\em ii)} It is not affected by completeness problems of the control 
field, thus saving telescope time. 
{\em iii)} The real radial extent of these stellar systems it is not known 
in advance. Therefore, the control fields might still be located inside their 
halo. The main drawback is that MW models need to be validated using deep and 
accurate star counts covering broad sky regions \citep[and references therein]{reyle09}.  

However, the ground-based and space data sets are characterized by different 
limiting magnitudes. To provide a robust estimate of the completeness of the former 
data sets, we compared their luminosity function (LF) with the LF of the WFPC2 data 
for pointing ``W". We adopted this approach since we are interested in estimating the 
completeness down to 1.5 magnitudes fainter than the TRGB, i.e., $I\lesssim 23$~mag.	    
In this magnitude range the WFPC2 data are minimally affected by completeness problems.  
We have found that the completeness is $\sim65$\% for the Subaru at the limiting magnitude  
of $I=23.0$ mag, while it is $\sim65$\% for the CFHT at the limiting magnitude of $I=22.6$ mag. 
We have chosen the above limiting magnitudes to apply conservative completeness corrections
to both the Suprime-Cam and the Mega-Cam data. 

In order to compare theory and observations, we used the Pisa \citep{castellani02,cignoni06} 
and the Padova \citep{girardi05} MW model. We focused our attention on two outer regions 
covering the same sky area, with the same mean reddening ($E(\bmv)=0.63\pm0.10$ mag) and 
located in the NE direction, namely the ``S4" (N(stars)$\sim$4000) and ``M2" 
(N(stars)$\sim$ 3000) regions. 
The top panels of Fig.~3 show the $I$,$\vmi$ CMD for the Pisa (left) and the
Padova (right) MW models for a sky area of $1^{\circ} \times 1^{\circ}$ at the
position of IC10 ($l=119^{\circ}$; $b=-3.3^{\circ}$) and assuming a reddening
of $E(\bmv)=0.63\pm0.10$ mag.

\begin{figure}[htbp]
\begin{center}
\includegraphics[height=.45\textheight,width=0.45\textwidth]{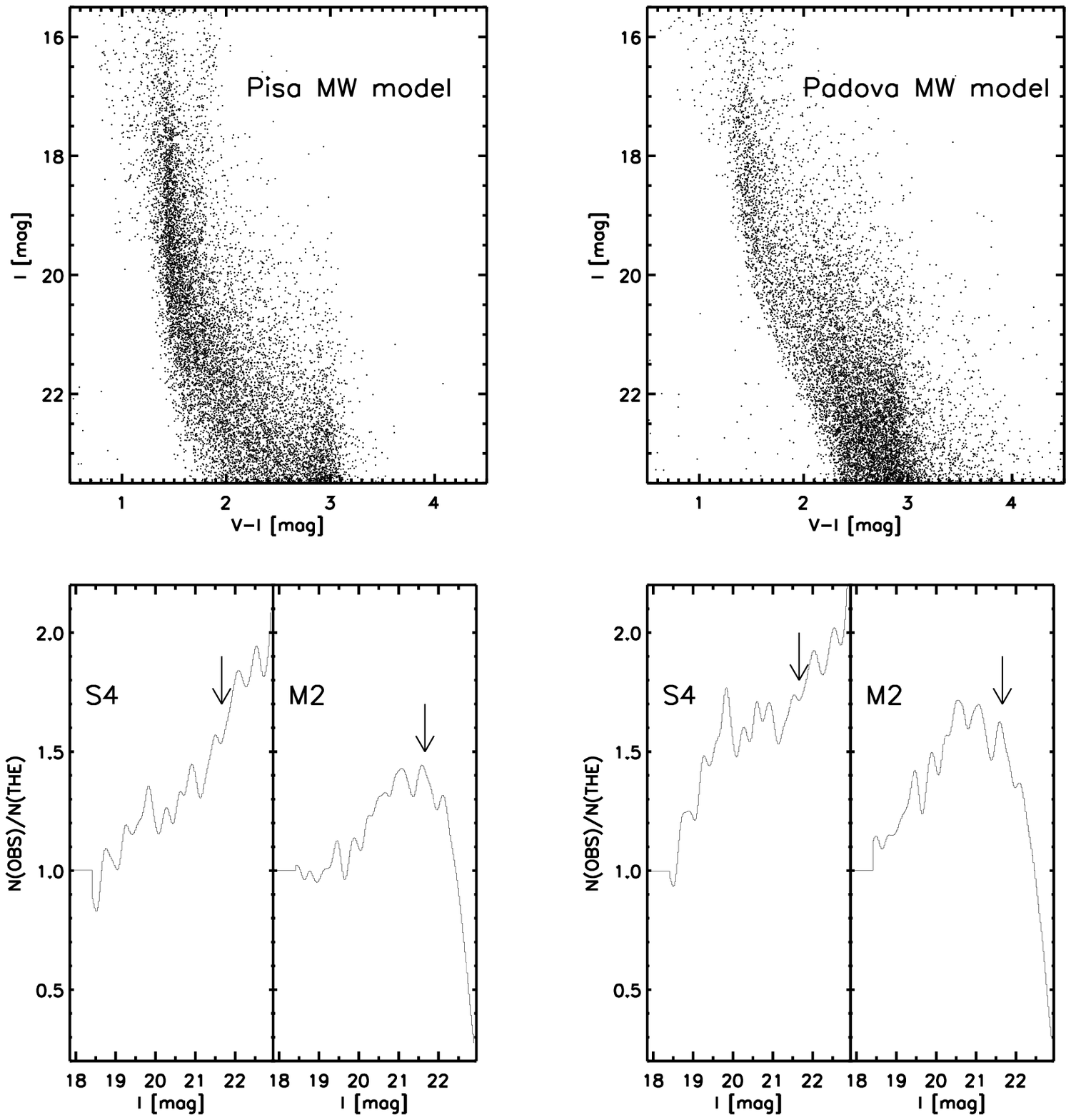}
\caption{Top--Simulated CMDs for the field stars adopting a reddening of
$E(B-V)=0.63\pm0.10$ mag in a field of view of $1^{\circ}$x $1^{\circ}$, according  
to the Pisa (left) and to the Padova (right) MW model. Only a small fraction of the 
total number of stars is plotted. 
Bottom--Left--Star count ratios between the stars located in the ``S4" and in the 
``M2" region with the Pisa MW model. There is evidence of IC10 stars at the position 
of the TRGB, i.e. $I=21.66\pm0.25$ mag (see the vertical arrows). 
Bottom--Right--Same as the left, but the ratio is between observations and the 
Padova MW model.}
\end{center}
\end{figure}

The bottom panels of Fig.~3 show the ratio between the number of observed stars 
and the number of candidate field stars predicted by the quoted MW models. Star 
counts were smoothed using a Gaussian kernel at fixed $\sigma$. We also 
estimated the number of unresolved background galaxies, 
with reddened $I$-band magnitudes and \vmi\  colors typical of the tip of 
RGB stars, 
using empirical galaxy counts \citep{fukugita95,benitez00,capak04,ferguson04} 
and it was subtracted to the number of observed stars. To avoid spurious fluctuations 
caused by the limited sky area covered by observations, theory and observations were 
normalized in the bright end (17.9$\le$I$\le$18.4 mag). The bottom left panels display the 
comparison with the Pisa MW model. There is evidence of IC10 stars across the 
TRGB region ($I=21.66 \pm 0.25$ mag) in both the ``S4" (left, 1.59$\pm$0.13) and the 
``M2" (right, 1.38$\pm$ 0.12) field. The star count excess is at 3$\sigma$ level. 
The above evidence further supports the occurrence of IC10 stars in the region covered 
by the Mega-Cam data, since in the ``S4" region we clearly identified IC10 RG stars.  
The bottom right panels of Fig.~3 display the comparison between observations and 
the Padova MW model. There is once again a clear evidence of a star excess across 
the TRGB region  in both the ``S4" (left, 1.74$\pm$0.14) and the 
``M2" (right, 1.53$\pm$0.14) field ($\sim$ 4$\sigma$ level). 
The two Galactic models were constructed assuming similar input parameters\footnote{Galactic 
model input parameters: Kroupa initial mass function not corrected for binaries; 
double exponential thin disk ($h_z$(height)=250 pc, $h_R$(scale length)=3000 pc, 
constant star formation rate [SFR] for t$\le$ 7 Gyr, Z(mean metallicity)=0.02); 
exponential thick disk ($h_z$=1000 pc, $h_R$=3500 pc, constant SFR for 5$\le$t$\le$12 Gyr, Z=0.006); 
oblate halo with $r^{1/4}$ ($h_R$=2800 pc, $h_q$(semiaxis ratio)=0.8, constant SFR for 
11$\le$t$\le$13 Gyr, Z=0.0002).}.    
The difference in the star count ratios between the two models is due to the different 
evolutionary inputs and to the normalization of the star counts in the Solar neighbourhood. 
 
The above findings indicate that the radial distribution of IC10 old and intermediate-age stellar 
populations agrees quite well with the size of the huge hydrogen cloud detected by \citet{huchtmeier79} 
and by \citet{cohen79}, and cover more than one square degree across the galaxy ($r\approx 34-42$\min). 
This means that the stellar halo and the hydrogen cloud have, within the errors, similar radial 
extents and resolve this peculiar feature of IC10 (\citealp{tikhonov09}). Moreover, this evidence 
further supports the hypothesis that the hydrogen cloud is associated with the galaxy 
\citep[stellar mass loss, pristine gas;][]{huchtmeier79,cohen79,wilcots98}.  

To estimate the total luminosity we need to select candidate IC10 stars. To describe 
the procedure, Fig.~4 shows the $I,\vmi$ CMDs of the stars located inside a circle of 13$'$ 
diameter across the galaxy center. The top panels display the photometry of space data 
(pointings $\alpha$, $\beta$, $\gamma$, $\delta$), while the bottom ones show
ground  data  (pointing $\epsilon$, Suprime-Cam). The difference between $\epsilon$1 
and $\epsilon$2 is in the mean reddening (see Fig.~4). For the stars located 
in the overlapping regions we use the HST photometry. The candidate IC10 stars were selected 
using different boxes in the aforementioned CMDs. The green box includes young MS stars 
($\vmi \sim 1.5$ mag), the cyan box the intermediate-age stars ($\vmi \sim 3$, 
$17.5 \lesssim I \lesssim 21.5$ mag), while the pink box includes old and intermediate-age RG stars
($3 \lesssim \vmi \lesssim6$, $I \sim 22$ mag). The position of the boxes in the four CMDs 
was shifted according to the local reddening (see Fig.~1). The limiting magnitude of the 
boxes is $I=23.0$ according to the completeness experiment. 

The dashed and the 
dashed-dotted lines plotted in the bottom right panel of Fig.~4, show two young 
scaled Solar abundance isochrones \citep{pietrinferni04} at fixed metallicity ([M/H]=-0.66 dex) 
and ages of $t=6$ and 200 Myr, while the solid red line shows the old $\alpha$-enhanced 
isochrone with the same total metallicity and an age of $t=13$ Gyr. 
These isochrones validate the position of the boxes we adopted to pinpoint the different 
subpopulations of IC10. The same approach was adopted to select candidate IC10 stars located 
between the blue circle of Fig.~1 and the outermost regions of the $\epsilon$ pointing 
($r \lesssim 23$\min).  
On the basis of these data and of the recent IC10 distance based on the
TRGB \citep[830 kpc,][]{sanna08} we estimated a total $V$-band luminosity
of $L_V\sim9.13\times 10^7$ $L_{\odot}$ and a total magnitude of $M_V$=-15.11 mag. 
This estimate agrees within a factor of two with similar estimates available in the 
literature ($L_V\sim1.6\times 10^8$ $L_{\odot}$, \citealt{mateo98}; $M_V$=-16.0 mag, 
\citealt{richer01}). 
The current estimate is a lower limit, since we are not including the stars located in the 
outermost regions covered by our photometry (pointing $\zeta$). However, the estimates 
available in the literature only cover the innermost galactic regions. 
The difference is mainly due to the fact that the current photometry allows us a robust identification 
of field stars ($1 \lesssim \vmi \lesssim 2$, $15 \lesssim I \lesssim 22$ mag, see the blue arrow in the 
bottom left panel of Fig.~4).	
If they are even partially included, these objects introduce a systematic bias in the estimate 
of the total luminosity. The different assumptions concerning the adopted distance and reddening 
variation also help to explain the above difference. 

\begin{figure*}[htbp]
\begin{center}
\includegraphics[height=.65\textheight,width=0.75\textwidth]{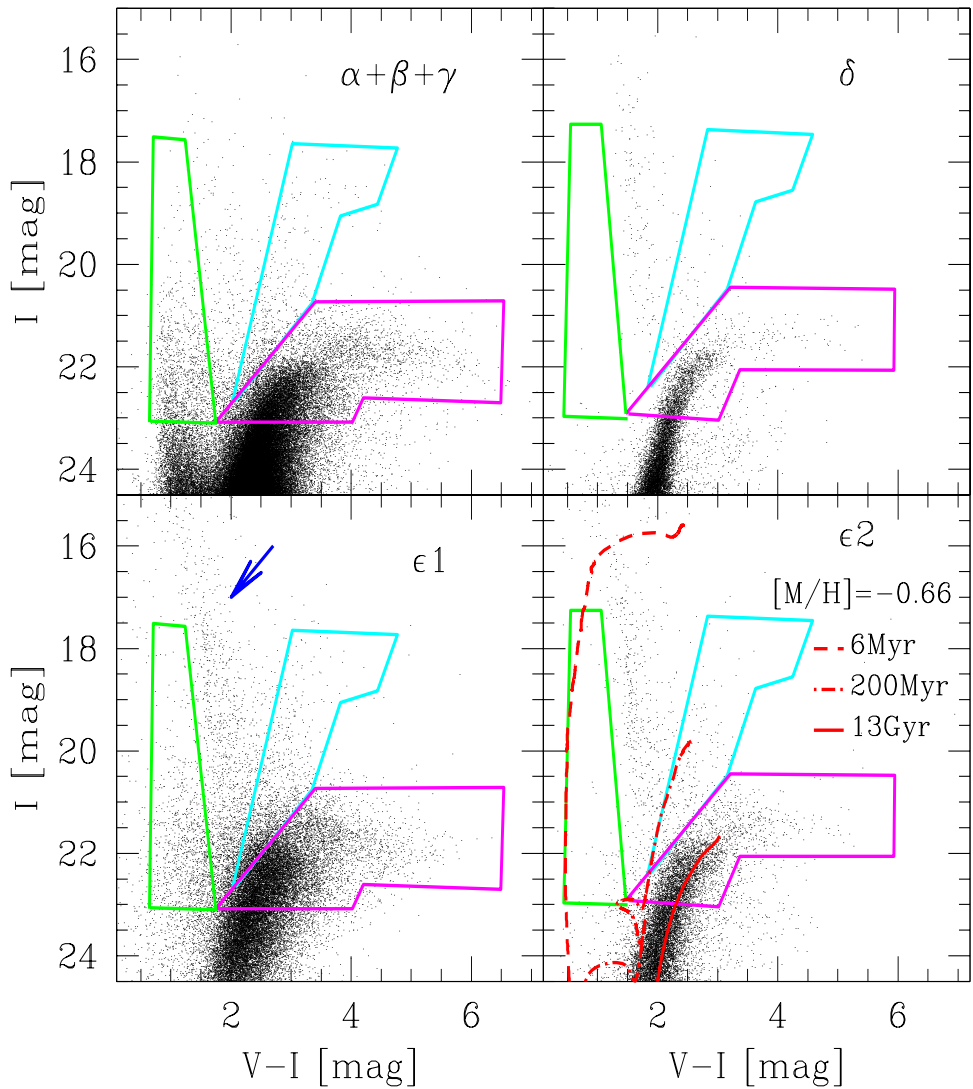}
\caption{CMDs in I, V-I bands for the stars located within $6.5'$ across the galaxy center.
Top -- space data of the pointings $\alpha$, $\beta$ and $\gamma$ (left, $E(B-V)=0.78\pm0.06$ mag) 
and $\delta$ (right, $E(B-V)=0.63\pm0.09$ mag).  
The colored boxes mark the CMD regions adopted to select candidate IC10 stars. The green box 
includes young MS stars, the cyan box the intermediate-age stars and the pink one the old and 
intermediate-age RGs. The position of the boxes was changed according to the local reddening. 
Bottom -- ground based data of the pointing $\epsilon$ ($\epsilon$1, $E(B-V)=0.78\pm0.10$; 
$\epsilon$2, $E(B-V)=0.63\pm0.10$ mag). The blue arrow marks the position of candidate field 
stars (left). The dashed and the dashed-dotted lines (right) display two young, scaled Solar 
isochrones (t=6, 200 Myr), while the solid line an old $\alpha$-enhanced isochrone (t=13 Gyr). 
Young and old isochrones were constructed at fixed total metallicity ([M/H]=-0.66 dex).}
\end{center}
\end{figure*}

To estimate the mass-to-light (M/L) ratio of IC10 we restricted ourself to
the galactic regions where rotational velocity measurements are available (see
the blue circle with a diameter of 13$'$ in Fig.~1). The luminosity inside this
area is $L_V\sim 5.88 \times10^7 L_{\odot}$ ($M_V=-14.63$~mag). By using the
quoted true distance and diameter together with the rotation velocity based on
H{\footnotesize{I}} regions (SS89) we found, following \citet{huchtmeier01} 
and \citet{case80}, a total mass of $M_{tot}\sim6.2$x$10^8M_{\odot}$ that 
agrees quite well with similar estimates available in the literature
\citep{huchtmeier79,vandenbergh00,woo08}. Eventually, we found M/L
$\sim$10 $M_{\odot}/L_{\odot}$. 
Although this estimate is hampered by several empirical limitations it
is at least one order of magnitude larger than the value recently provided
by \citet{woo08}. The quoted authors use two independent methods to estimate
the M/L ratios of LG dwarf galaxies: colors and inferred star formation
history.  They found that the median M/L ratio of dIs based on the latter
approach is slightly smaller than on the former one (0.7 vs 0.8, see their
Table˜2).   However, the difference needs to be investigated in more detail,
particularly in view of the severe limitations affecting the estimates of the
rotational velocity and of the total luminosity over the entire body of the
galaxy. 

We are facing empirical evidence that dIs seem to show smaller M/L ratios
when compared with dwarf ellipticals (see Fig.~1 and Table~2 in \citealt{woo08}). 
The new data will allow us to constrain whether this evidence might be affected 
by observational biases. 
Moreover, they can shed new lights on the prediction that dwarf galaxies 
might have tidal radii significantly larger than empirical estimates 
\citep{hayashi03,kazantzidis04}. 

\acknowledgments
It is a real pleasure to thank the referee, Dr. I. Chilingarian, for his constructive  
suggestions.



\begin{thebibliography}{}

\bibitem[Bekki (2008)]{bekki08} Bekki, K. 2008, MNRAS, 388,10

\bibitem[Benitez (2000)]{benitez00} Benitez, N. 2000, ApJ, 536, 571

\bibitem[Capak et al. (2004)]{capak04} Capak, P. et al. 2004, AJ, 127, 180

\bibitem[Casertano \& Shostak (1980)]{case80} Casertano, S. P. R., \& Shostak, G. S.  1980, A\&A, 81, 371

\bibitem[Castellani et al. (2002)]{castellani02} Castellani, V., Cignoni, M., Degl'Innocenti, S., Petroni, S., Prada Moroni, P. G. 2002, MNRAS, 334, 69

\bibitem[Cignoni et al. (2006)]{cignoni06} Cignoni, M., Degl'Innocenti, S., Prada Moroni, P. G., Shore, S. N. 2006, A\&A, 459, 783

\bibitem[Cohen (1979)]{cohen79} Cohen, R. J. 1979, MNRAS, 187, 839

\bibitem[Demers et al. (2004)]{demers04} Demers, S., Battinelli, P., Letarte, B. 2004, A\&A, 424, 125

\bibitem[Ferguson et al. (2004)]{ferguson04} Ferguson, H. C. et al.\ 2004, ApJ, 600L, 107

\bibitem[Fukugita et al. (1995)]{fukugita95} Fukugita, M., Shimasaku, K., Ichikawa, T. 1995, PASP, 107, 945

\bibitem[Girardi et al. (2005)]{girardi05} Girardi, L., Groenewegen, M. A. T., Hatziminaoglou, E., da Costa, L 2005, A\&A, 436, 895


\bibitem[Hayashi et al. (2003)]{hayashi03} Hayashi, E., Navarro, J. F., Taylor, J. E., Stadel, J., Quinn, T. 2003, ApJ, 584, 541

\bibitem[Huchtmeier (1979)]{huchtmeier79} Huchtmeier, W. K. 1979, A\&A , 75, 170 (H79)

\bibitem[Huchtmeier \& Richter (2001)]{huchtmeier01} Huchtmeier, W. K., Richter, O. -G. 1988, A\&A , 203, 237

\bibitem[Hunter (2001)]{hunter01} Hunter, D. A. 2001, ApJ, 559, 225

\bibitem[Jarrett et al. (2003)]{jarrett03} Jarrett, T. H. Chester, T., Cutri, R., Schneider, S. E., Huchra, J. P. 2003, AJ, 125, 525

 \bibitem[Kazantzidis et al. (2004)]{kazantzidis04} Kazantzidis, S., Mayer, L., Mastopietro, C., Diemand, J., Stadel, J., Moore, B. 2004, ApJ, 608, 663

\bibitem[Kormendy et al.(2009)]{kormendy09} Kormendy, J., Fisher, D.~B., Cornell, M.~E., \& Bender, R.\ 2009, \apjs, 182, 216


\bibitem[Landolt (1983)]{landolt83} Landolt, A.U., 1983, AJ, 88, 853

\bibitem[Landolt (1992)]{landolt92} Landolt, A. U. 1992, AJ, 104, 340

\bibitem[Massey \& Armandroff (1995)]{massey95} Massey, P. \& Armandroff, T. E. 1995, AJ, 109, 2470

\bibitem[Mateo (1998)]{mateo98} Mateo, M. 1998, ARA\&A, 36, 435

\bibitem[McConnachie et al. (2008)]{mcconnachie08} McConnachie, A. W., Huxor, A., Martin, N. F. et al.\ 2008, ApJ, 688, 1009

\bibitem[Pietrinferni et al. (2006)]{pietrinferni06} Pietrinferni, A., Cassisi, S., Salaris, M., Castelli, F. 2006, ApJ, 642, 797

\bibitem[Pietrinferni et al. (2004)]{pietrinferni04} Pietrinferni, A., Cassisi, S., Salaris, M., Castelli, F. 2004, ApJ, 612, 168

\bibitem[Reyle et al. (2009)]{reyle09} Reyle, C., Marshall, D. J., Robin, A. C., Schultheisé, M. 2009, A\&A, 495, 819

\bibitem[Richer et al. (2001)]{richer01} Richer, M. G. et al. 2001, A\&A, 370, 34

\bibitem[Sakai et al. (1999)]{sakai99} Sakai, S., Madore, B. F., Freedman, W. L. 1999, ApJ, 511, 671


\bibitem[Sanna et al.(2008)]{sanna08} Sanna, N. et al. 2008, ApJ, 688, L69

\bibitem[Sanna et al.(2009)]{sanna09} Sanna, N. et al. 2009, ApJ, 699, L84

\bibitem[Shostak \& Skillman (1989)]{shostak89} Shostak, G. S., Skillman, E. D. 1989, 1989, A\&A, 214, 33 (SS89)

\bibitem[Sirianni et al. (2005)]{sirianni05} Sirianni, M. et al. 2005, PASP, 117, 1049

\bibitem[Stetson (1994)]{stetson94} Stetson, P. B. 1994, PASP, 106, 250

\bibitem[Stetson (1987)]{stetson87} Stetson, P.B, 1987, PASP, 99, 191

\bibitem[Tikhonov \& Galazutdinova (2009)]{tikhonov09} Tikhonov, N. A., Galazutdinova, O. A. 2009, AstL, 35, 748

\bibitem[Tolstoy et al.(2009)]{tolstoy09} Tolstoy, E., Hill, V., \& Tosi, M.\ 2009, \araa, 47, 371

\bibitem[Vacca et al. (2007)]{vacca07} Vacca, W. D.; Sheehy, C. D., Graham, J. R. 2007, ApJ, 662, 272


\bibitem[van den Bergh (2000)]{vandenbergh00} van den Bergh, S. 2000, The Galaxies of the Local Group, ed. Cambridge Astrophysical Series

\bibitem[Wang et al. (2005)]{wang05} Wang, Q. D., Whitalker, K. E., Williams, R. 2005, MNRAS, 362, 1065

\bibitem[Wilcots \& Miller (1998)]{wilcots98} Wilcots, E. M., Miller, B. W. 1998, AJ, 116, 2363

\bibitem[Woo et al. (2008)]{woo08} Woo, J., Courteau, S., Dekel, A. 2008, MNRAS, 390, 1453

\bibitem[Wyse (2010)]{wyse10} Wyse, R. F. G. 2010, AN, 331, 526


\end{thebibliography}
\end{document}